\documentclass[prx,twocolumn]{revtex4-1}
\usepackage{natbib}
\usepackage{color}
\usepackage{array}
\usepackage{graphicx}
\definecolor{myblue}{RGB}{0, 50, 100}
\usepackage[margin=0.7in]{geometry}
\usepackage{amsmath}
\usepackage{amssymb,bm}
\usepackage{amsthm}
\usepackage{mathtools}
\usepackage{url}
\usepackage{mathrsfs}
\usepackage{mathtools}
\usepackage{float}
\usepackage{verbatim}
\usepackage{latexsym}
\usepackage{setspace}
\usepackage{amsfonts}
\usepackage{stmaryrd}
\usepackage{xcolor}
\usepackage{enumitem}
\usepackage{cleveref}
\usepackage{braket}
\usepackage{siunitx}

\date{September 16, 2020}

\begin{document}

\title{Laser-annealing Josephson junctions for yielding scaled-up superconducting quantum processors.}

\author{Jared B. Hertzberg}
\author{Eric J. Zhang}
\author{Sami Rosenblatt}
\author{Easwar Magesan}
\author{John A. Smolin}
\author{Jeng-Bang Yau}
\author{Vivekananda P. Adiga}
\author{Martin Sandberg}
\author{Markus Brink}
\author{Jerry M. Chow}
\author{Jason S. Orcutt}
\affiliation{IBM Quantum, IBM T.J. Watson Research Center, Yorktown Heights, NY 10598, USA}

\begin{abstract}
\noindent
As superconducting quantum circuits scale to larger sizes, the problem of frequency crowding proves a formidable task. Here we present a solution for this problem in fixed-frequency qubit architectures. By systematically adjusting qubit frequencies post-fabrication, we show a nearly ten-fold improvement in the precision of setting qubit frequencies. To assess scalability, we identify the types of `frequency collisions' that will impair a transmon qubit and cross-resonance gate architecture. Using statistical modeling, we compute the probability of evading all such conditions, as a function of qubit frequency precision. We find that without post-fabrication tuning, the probability of finding a workable lattice quickly approaches 0. However with the demonstrated precisions it is possible to find collision-free lattices with favorable yield. These techniques and models are currently employed in available quantum systems and will be indispensable as systems continue to scale to larger sizes.
\end{abstract}

\maketitle{}

\section{Introduction} \label{sec:intro}

Realizing robust large-scale quantum information processors  is one of the foremost challenges in quantum science. Many practical applications have been proposed for robust quantum computers, including estimating the ground state energy of chemical compounds and implementing machine learning algorithms \cite{kandala_hardware-efficient_2017, havlicek_supervised_2019, hempel_quantum_2018, colless_computation_2018, nam_ground-state_2020}. Quantum advantage relative to classical computers can be realized without full fault-tolerance, but requires large quantum circuits that a classical computer cannot simulate \cite{kandala_error_2019}. Recent demonstrations have shown qubit circuits nearly at the threshold for demonstrating quantum advantage \cite{corcoles_challenges_2019}. Much work remains in order to realize fault-tolerant quantum processors; however,  scale-up of solid-state quantum circuits has shown consistent and ongoing progress \cite{chow_implementing_2014, corcoles_demonstration_2015, gambetta_building_2017, takita_demonstration_2016, takita_experimental_2017, kelly_state_2015, riste_detecting_2015, ofek_extending_2016}. As the qubit circuits are scaled up, they must maintain high one- and two-qubit gate fidelities, high qubit connectivity, and low cross-talk error which can be measured in a holistic sense via the quantum volume of the circuit \cite{cross_validating_2019}. Lattices of fixed-frequency transmon qubits represent a promising architecture for building systems of larger sizes \cite{corcoles_challenges_2019}. A growing number of systems at the 20 to 50-qubit scale are now available to users through cloud access. A variety of technical challenges confront further system scaling, including improving 3-dimensional circuit integration and qubit coherence. High on the list of such challenges is the issue of `frequency crowding.'\\

Fixed frequency transmon qubits using the two-qubit cross-resonance (CR) gate form a promising architecture for scaling up quantum systems. Fixed-frequency transmons are largely insensitive to charge or flux noise, and have achieved coherence times of 100 $\mu$s and growing. The CR gate, a hardware-efficient all-microwave gate \cite{chow_simple_2011, chow_universal_2012, de_groot_selective_2012, rigetti_fully_2010}, is readily used to entangle these qubits with gate fidelities above 99\%, approaching the threshold for fault-tolerant codes \cite{sheldon_procedure_2016}. To achieve these fidelities, the CR gate needs not only high coherence qubits, but also a precise setting of the qubits' frequencies. The CR gate activates a $ZX$ interaction by driving one `control' qubit with a microwave pulse at the other `target' qubit's transition frequency. The magnitude of the $ZX$ as well as other Hamiltonian terms depends on the relative frequencies of the two qubits \cite{magesan_effective_2020, ware_cross-resonance_2019}. Diminished $ZX$ magnitude increases gate time, while other terms such as $ZZ$ add gate errors. Neighboring qubits having the wrong detuning will exhibit a `frequency collision' in which the $ZX$ may be suppressed or other undesirable effects arise. \\

Maintaining high gate fidelities for all pairs in a lattice will require solving this `frequency crowding' problem by precise setting of qubit frequencies to specified values, as characterized by a standard deviation $\sigma_f$. To achieve low $\sigma_f$, the tunnel-junction conductance must be controlled with high precision. Transmon frequency $f_{01}$ follows $h f_{01} \simeq \sqrt{8 E_J E_C} - E_C$, where Josephson energy $E_J = \frac{\hbar I_c}{2 e}$ is many times greater than charging energy $E_C = \frac{e^2}{2 C}$ \cite{koch_charge-insensitive_2007}. In typical transmons, a photolithographically defined capacitance $C$ has dimensions in the tens to hundreds of microns and varies little from qubit to qubit. The critical current $I_c$ is set by a tunnel barrier of area $\sim$ 100 $\times$ 100 nm and thickness a few nm, and is thus challenging to fabricate with precision better than a few percent \cite{potts_cmos_2001,wu_overlap_2017,costache_lateral_2012,kreikebaum_improving_2020}. However, tunnel barrier resistance $R_n$ is readily measurable to precision better than 0.1\% and relates to $I_c$ according to the Ambegaokar-Baratoff relation $I_c = \frac{\pi \Delta}{2 e R_n}$ (where $\Delta$ is the superconducting gap energy) \cite{ambegaokar_tunneling_1963}. We can therefore measure $R_n$ before a chip is cooled in order to assess qubit frequency imprecision. The best demonstrated precision in setting $R_n$ at time of fabrication is 2\% \cite{kreikebaum_improving_2020}.  A 2\% variation in $R_n$ indicates a fractional $\sigma_f$ of 1\%. \\

Careful design of lattices can enable error correction codes while at the same time minimizing the likelihood of `frequency collisions' and therefore the required $\sigma_f$ for fabrication yield \cite{chamberland_topological_2020, chow_reducing_2020}. Yet even the most robust designs require a fractional $\sigma_f$  of 0.25\% to 0.5\%, which represents a factor of 2 to 4 improvement over the best literature results. To overcome such limits will require rework of individual qubits' tunnel junctions after fabrication. Thermal anneal has been shown to increase tunnel resistance $R_n$, and laser heating has been demonstrated as a highly localized re-work tool \cite{koppinen_complete_2007, granata_trimming_2008, muthusubramanian_b29015_2019, oliva_annealing_1994, lehnert_thermal_1992, orcutt_united_2019}. However, the inherent variability of the anneal process itself must be overcome, and qubit frequency control utilizing such techniques at scale has never been presented in the literature.\\

In this paper, we introduce an adaptive post-fabrication trimming technique that we use to incrementally adjust $R_n$ on a qubit-by-qubit basis, thereby overcoming inherent variability in both initial qubit fabrication and the laser anneal. For the first time, an improvement in qubit frequency precision is demonstrated in terms of narrowed frequency distributions. Crucially, we demonstrate qubit frequency imprecision $\sigma_f$ of the same magnitude as the imprecision of predicting $f_{01}$ from $R_n$. To estimate the scalability of this technique for the fabrication of error-corrected lattices, we employ a statistical yield model based on $\sigma_f$ relative to specific collision bounds. This model predicts the severity of the frequency crowding problem for different topologies and scales of error corrected multi-qubit lattices as a function of code distance. The model demonstrates that using conventional transmon fabrication, scaled-up qubit lattices will fail to evade `frequency collisions'. However, our novel trimming technique achieves adequate $\sigma_f$ for scalable fabrication of distance-3 through distance-7 heavy-square and heavy-hexagon codes. In particular, this technique enables the high yield fabrication of the distance-3 and distance-5 heavy-hexagon lattices currently deployed as IBM cloud connected systems \cite{jurcevic_demonstration_2020}.

\section{Results} \label{sec:results}

\subsection{Frequency Precision $\sigma_f$ From Transmon Fabrication}

\begin{figure}
	\includegraphics[width=0.44\textwidth, trim=0.0cm 0.0cm 0.0cm 0.0cm, clip]{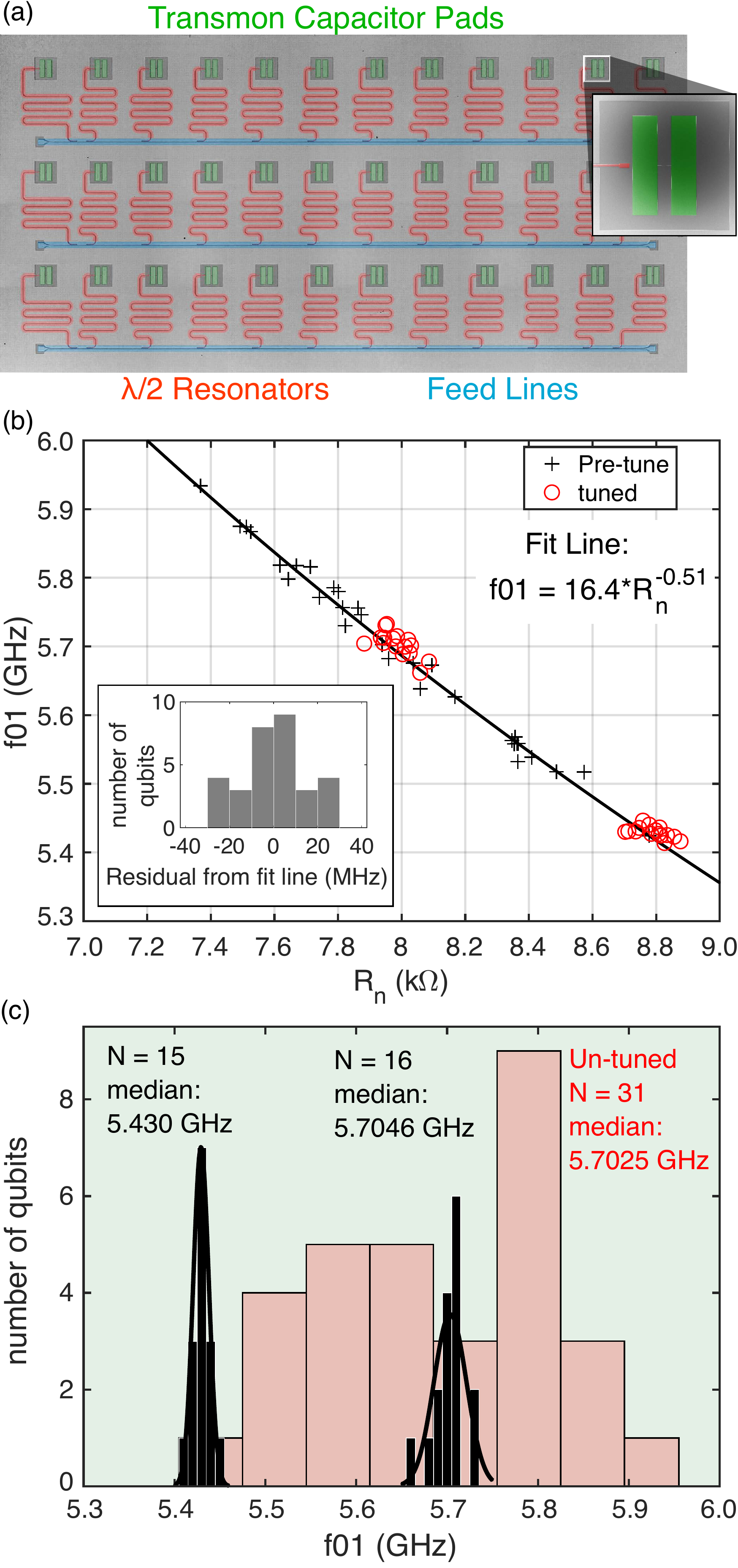}
	\caption{\textbf{(a) Chip type used to assess $\sigma_f$.} False-colored image. 36 fixed-frequency transmon qubits, each including a 500 $\times$ 320 micron planar capacitor and a $\sim 0.1 \times 0.1$ micron Al/AlOx/Al tunnel junction, are prepared identically on a 20 $\times$ 10 mm Si substrate. A half-wavelength coplanar waveguide resonator at each qubit enables dispersive readout. Resonators are frequency-multiplexed in groups of 12. \textbf{(b) Transmon frequency vs $R_n$.} Power law fit of pre-tuned population. Inset histogram (10 MHz bins) shows residual scatter in frequency relative to fit line. \textbf{(c) Distributions of qubit frequencies.} Initial median was 5.7025 GHz and spread $\sigma_f = 132.3$ MHz (red histogram, 70 MHz bins). Using selective laser-anneal (figure \ref{fig:LaserTuneSchematic}) we prepared these qubits into two distinct frequency populations with medians 5.430 and 5.7046 GHz. (Black histograms, 10 MHz bins) Each population is outlined by a gaussian curve centered at its local median frequency. Combined spread is $\sigma_f = 14.0$ MHz.} \label{fig:36Q_andTuned2f}
\end{figure}

To assess the $\sigma_f$ resulting from qubit fabrication, we developed a test vehicle containing a large number of identically-fabricated qubits (\cref{fig:36Q_andTuned2f}). We cooled the chip in a dilution refrigerator and used dispersive readout through half-wave microwave resonators to measure qubit frequencies \cite{blais_cavity_2004}. We measured the frequencies of 31 qubits to a precision better than 100 kHz using a Ramsey fringe method. The qubit frequencies had random variation $\sigma_f = 132.3$ MHz (\cref{fig:36Q_andTuned2f}), or 2.3\% of the median frequency. After warming the qubits to room temperature, we measured their junction resistances. In \cref{fig:36Q_andTuned2f}, we show a plot of $R_n$ compared to transmon frequency, demonstrating that the observed variation in $\sigma_f$ is accounted for almost entirely by $R_n$ variation. The behavior may be fit to a power law of approximately $-\frac{1}{2}$ power, as expected from transmon theory and the Ambegaokar-Baratoff formula. For a population of transmon qubits whose frequency scatter is dominated by scatter in $R_n$, we expect the fractional standard deviation in $R_n$ to be twice that of $\sigma_f$. This is consistent with the standard deviation in junction resistances which is found to be $\sigma_R$ of 365 $\Omega$, or 4.6\% of the median $R_n$. We also assess the fidelity of the frequencies to the $f$-vs-$R_n$ correlation in terms of the residual scatter after subtracting the fit line from the frequency values. This appears in the inset in \cref{fig:36Q_andTuned2f} and exhibits a standard deviation 14.5 MHz, or 0.25\% of the qubit median frequency.\\

\subsection{Tuning Using Selective Laser Anneal}

\begin{figure*}
	\centering
	\includegraphics[width=0.96\textwidth, trim=0.0cm 0.0cm 0.0cm 0.0cm, clip]{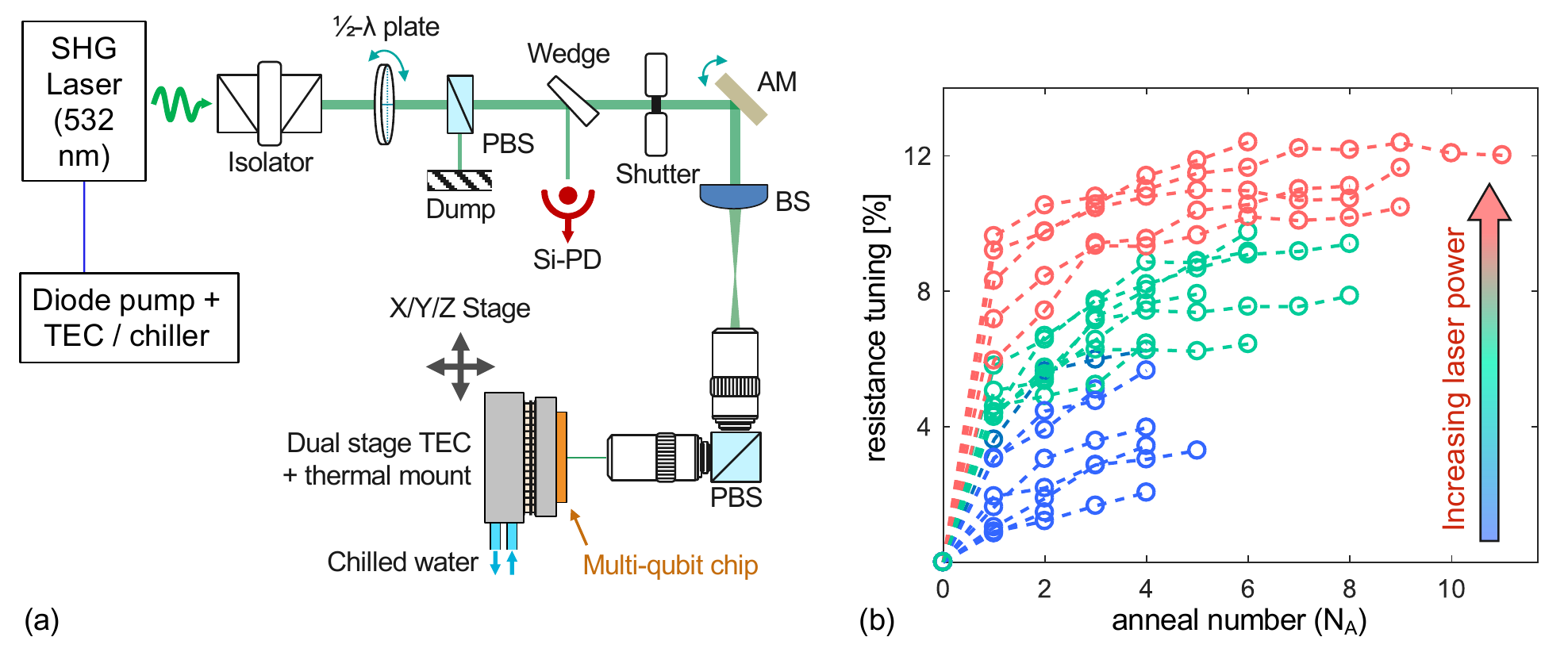}
	\caption{\textbf{(a) Schematic of apparatus, ``Laser Annealing of Stochastically Impaired Qubits'' (LASIQ).} A 532 nm (frequency doubled) diode-pumped solid-state laser is used as the laser annealing source. Active power calibration is accomplished via a half-wave plate and PBS combination, with feedback from a Si-PD. A piezoelectric mirror mount actively aligns the beam to the junction center via image pattern recognition . The beam is shaped as necessary to avoid direct illumination of the junction, and beam size is condensed $4\times$ using a dual-objective setup \cite{rosenblatt_laser_2019-1}. \textbf{(b) Iterative anneal demonstration}. Adaptive anneal progression towards $R_n$ targets in 20 tunnel junctions. Greater fractional tuning requires greater anneal powers and anneal numbers. Abbreviations: SHG: second harmonic generation, PBS: polarizing beam splitter, Si-PD: silicon photodiode, AM: alignment mirror, BS: beam shaper, TEC: thermoelectric cooler.} \label{fig:LaserTuneSchematic}
\end{figure*}

To reduce $\sigma_f$, we developed a technique for selective laser anneal to shift tunnel resistance $R_n$ by pre-calibrated increments. (See \cref{sec:methods} and \cref{fig:LaserTuneSchematic}). We demonstrate the achievable frequency control of this technique by shifting the 31 measured qubits into a two-frequency pattern. We employed an $R_n$ vs $f$ correlation (\cref{fig:36Q_andTuned2f}) to designate the target resistances. We shifted 16 junctions to one target $R_n$ and 15 to another target $R_n$. After tuning, the group of 16 junctions had median resistance 7.984 k$\Omega$ and the group of 15 had median resistance 8.798 k$\Omega$. The 31 junctions reached their targets with an overall precision of $\sigma_R = 51 \Omega$, about 0.61\%. In a dilution refrigerator, we re-measured the frequencies of the qubits in the two groups. Aside from two of the qubits, which we measured using CW spectroscopy (precision 2 MHz), all qubits were remeasured in the same way as in the first cooldown. The resulting frequencies appear in \cref{fig:36Q_andTuned2f}. The two frequency groups are approximately normally-distributed and have medians $f_{0,1} = 5.430$ GHz and $f_{0,2} = 5.7046$ GHz. Calculating $\sigma_f = \sqrt{\left<(f_i - f_{0,j})^2\right>}$, where $f_{0,j}$ represents $f_{0,1}$ or $f_{0,2}$ as appropriate for a given qubit $Q_i$, we assess the overall precision $\sigma_f = 14.0$ MHz. This imprecision is nearly identical to the residual scatter from the $f(R)$ fit line (\cref{fig:36Q_andTuned2f}) which guided the tuning, and the fractional precision $\sigma_f / \left< f \right> = $ 0.25\% is slightly better than half of the fractional precision in setting $R_n$. By these comparisons, we see that in this experiment $\sigma_f$ is limited by both the precision of setting $R_n$ and the precision of predicting $f$ from $R_n$. Drift in $R_n$ reported in the literature \cite{koppinen_complete_2007} does not appear to be a limiting factor in this study. Achieving smaller $\sigma_f$ will require improvements in setting $R_n$. As we show in \cref{sec:methods}, the laser-anneal tuning technique is capable of precisions of 0.3\% in $R_n$. On that basis the imprecision of 14.5 MHz in predicting $f$ from $R_n$ would dominate the imprecision in $\sigma_f$.

\section{Discussion} \label{sec:discuss}

\begin{figure*}
	\includegraphics[width=0.96 \textwidth, trim=0.0cm 0.0cm 0.0cm 0.0cm, clip]{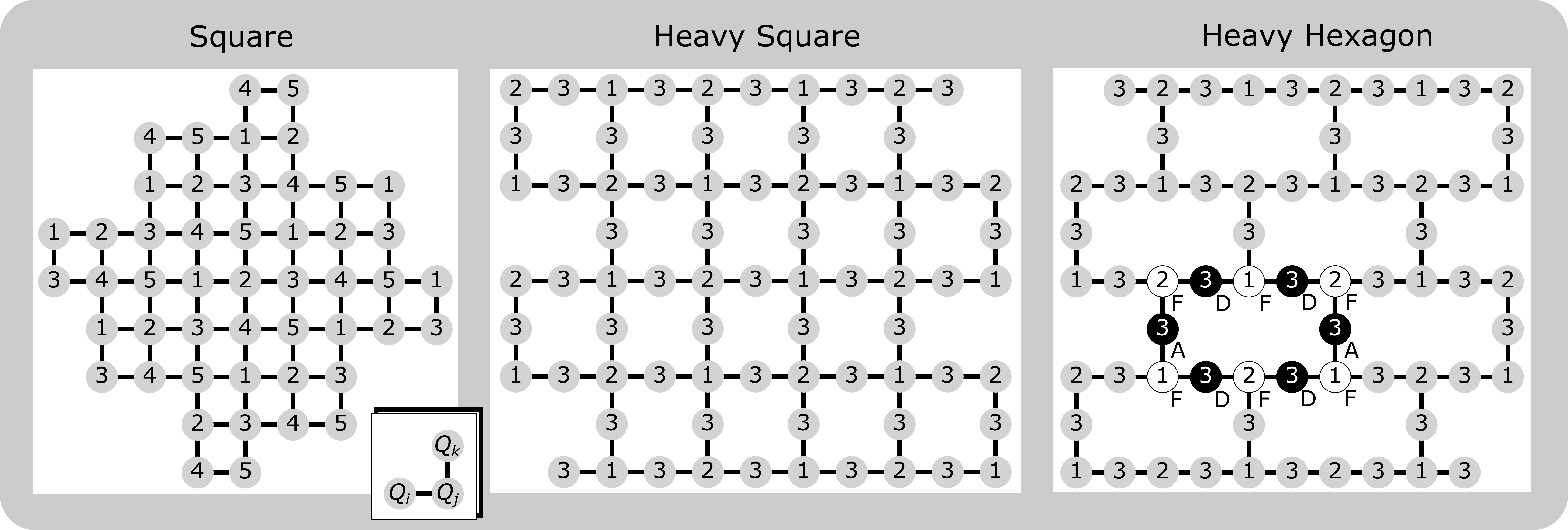}
	\caption{\textbf{Lattice and frequency-pattern examples.} Lattices are capable of $d=5$ codes. Patterns of qubit frequencies avoid all conditions in \cref{table:collision_types}. Statistical model is applied to these examples and to equivalent lattices at $d = 3$ and $d = 7$. (See Supplementary Figures 1-9.) Square lattice includes 49 qubits in 5-frequency pattern. Heavy-square lattice includes 73 qubits in 3-frequency pattern. Heavy-hexagon lattice includes 65 qubits in 3-frequency pattern. In a portion of the heavy-hexagon lattice, we indicate qubits' intended gate roles: control (black circles) or target (white circles); as well as code roles: data (D), ancilla (A) or flag (F) \cite{chamberland_topological_2020}. Inset shows relative positions of qubits for collision definitions of \cref{table:collision_types}. $Q_j$ is coupled to nearest-neighbors $Q_i$ and to $Q_k$. Qubits $Q_i$ and $Q_k$ are next-nearest-neighbors.}
	\label{fig:lattices_and_freqs}
\end{figure*}

Our post-fabrication trimming reduced $\sigma_f$ by 9.5$\times$ compared to initial fabrication. To assess whether this level of precision is sufficient to reliably prepare lattices of fixed-frequency transmons capable of error-correcting codes, we must quantify the frequency-crowding problem. Transmon qubits are weakly anharmonic and have decreasing transition energies at higher levels. Therefore, degeneracies among the $\ket{0} \rightarrow \ket{1}$, $\ket{1} \rightarrow \ket{2}$ and $\ket{0} \rightarrow \ket{2}$ transitions of nearby qubits can all contribute to `frequency collisions.' We must consider the relative frequencies of both nearest-neighbors and next-nearest-neighbors in the lattice \cite{magesan_effective_2020, mckay_three-qubit_2019, malekakhlagh_first-principles_2020}. Fig \ref{fig:lattices_and_freqs} illustrates the relative positions of nearest-neighbor and next-nearest-neighbor qubits in a section of lattice, and \cref{table:collision_types} lists the seven cases most likely to lead to gate errors \cite{magesan_effective_2020}. We can think of them qualitatively as follows: Type 1 causes hybridization of states in $Q_j$ and $Q_k$, while in type 2 the CR pulse excites $Q_j$ into the non-computational $\ket{2}$ state. Type 3 excites $Q_k$ to the $\ket{2}$ state, but does not require a CR tone. In condition 4, $ZX$ is weak, which implies long gate times and increased gate error \cite{ware_cross-resonance_2019, magesan_effective_2020}. In type 5, the CR gate addresses an additional neighboring qubit. In type 6, when one qubit is the target of a CR gate, its next-nearest-neighbor leaks to the $\ket{2}$ state. Type 7 causes $Q_j$ to leak to the $\ket{2}$ state during a CR gate.\\

\begin{table*}[t!]
	\centering
	\begin{tabular}{|c|c|c|c|} 
		\hline
		Type & Definition & Participants & Bounds\\ [0.5ex] 
		\hline
		1 & $f_{j,01} = f_{k,01}$ & Nearest-neighbor qubits $Q_j$, $Q_k$ & $\pm$ 17 MHz\\ 
		2 & $f_{j,02} = 2f_{k,01}$ & Control qubit $Q_j$, target qubit $Q_k$ & $\pm$ 4 MHz\\
		3 & $f_{j,01} = f_{k,12}$ & Nearest-neighbor qubits $Q_j$, $Q_k$ & $\pm$ 30 MHz\\
		4 & $f_{k,01} < f_{j,12}$ or $f_{j,01} < f_{k,01}$ &  Control qubit $Q_j$, target qubit $Q_k$ & --- \\
		5 & $f_{i,01} = f_{k,01}$ & $Q_j$ is control to $Q_i$ and/or $Q_k$ \& is nearest-neighbor to both. & $\pm$ 17 MHz\\
		6 & $f_{i,01} = f_{k,12}$ or $f_{i,12} = f_{k,01}$ & $Q_j$ is control to $Q_i$ and/or $Q_k$ \& is nearest-neighbor to both. & $\pm$ 25 MHz\\  
		7 & $f_{j,02} = f_{k,01} + f_{i,01}$ & $Q_j$ is control to $Q_i$ and/or $Q_k$ \& is nearest-neighbor to both. & $\pm$ 17 MHz \\
		\hline
	\end{tabular}
	\caption{\label{table:collision_types} \textbf{Seven most likely types of `frequency collision'.} For bus-coupled transmon qubits employing cross-resonance gates and having anharmonicity $\sim - 330$ MHz. Relative qubit positions illustrated in inset of \cref{fig:lattices_and_freqs}. Bounds for types 1,2,3 are estimated from model results \cite{magesan_effective_2020} as the region where gate errors due to `frequency collision' exceed $\sim$ 1 \%. Bounds for types 5,6,7 are based on those in types 1 and 3. }
\end{table*}

Around each of the `frequency collisions' described in  \cref{table:collision_types}, we can designate a window of undesired frequencies. This breaks the frequency space into allowed and forbidden regions. Type 4 listed in \cref{table:collision_types} defines forbidden zones where $ZX$ coupling is too low. For the other six conditions, we forbid regions where the `frequency collision' is the dominant source of gate error. Existing multi-qubit systems with CR gates typically exhibit two-qubit gate errors of 1 to 2 \% regardless of frequency \cite{takita_experimental_2017, mckay_three-qubit_2019}. Ref \cite{magesan_effective_2020} considers an effective-Hamiltonian model for the CR gate, as a function of the relative frequency of control and target qubits. We use this model to estimate the frequency windows for nearest-neighbor collisions (\cref{table:collision_types}, types 1 to 3) to cause gate errors exceeding $\sim$ 1 \%. We make an assumption that similar bounds apply to next-nearest-neighbor interactions (types 5 to 7).\\

A useful lattice of qubits should enable high quantum volume and fault-tolerant operation while avoiding all of the `frequency collisions' and forbidden regions presented in \cref{table:collision_types}. Both lattice layout and the pattern of qubit frequencies are relevant. We consider three types of lattices: square, `heavy-square' and `heavy-hexagon' (\cref{fig:lattices_and_freqs}). Lattices comprise qubits and two-qubit connections, each qubit being linked to no more than four neighbors. In many practical implementations, these links comprise microwave-resonant buses. A square lattice facilitates `surface code' fault-tolerant codes \cite{fowler_surface_2012}. Recent literature describes hybrids of the surface code with Bacon-Shor type codes, which can be employed in `heavy hexagon' and `heavy square' lattices to achieve fault-tolerance, albeit with lower error thresholds than the surface code \cite{chamberland_topological_2020}. In addition to the data and ancilla qubit roles employed in the surface code, these hybrid codes assign a portion of the lattice as `flag' qubits.\\

In the square lattice, every qubit in the bulk of the lattice lies on a degree-four vertex, while some at edges have degree two or degree three. If we populate the square lattice with 5 distinct frequencies of qubits, $f_5 > f_4 > f_3 > f_2 > f_1$, with appropriate spacing between the frequencies, we can avoid all the forbidden regions of  \cref{table:collision_types} \cite{gambetta_building_2017}. In \cref{fig:lattices_and_freqs}, we illustrate this pattern for square lattices capable of distance-5 ($d = 5$) rotated surface codes. Condition 4 of \cref{table:collision_types} requires $f_{control} > f_{target}$, so the pattern also fixes the direction of CNOT gate for each pair. \\

In contrast to the square lattice, the `heavy square' lattice includes both degree-two and degree-four vertices in the bulk. Degree-one, -two or -three vertices appear at the edges. We take advantage of this pattern to make all the degree-two vertices control qubits, using a three-frequency pattern $f_3 > f_2 > f_1$. Since every control qubit (frequency $f_3$) is linked to at most two target qubits, we need only two properly-chosen target-qubit frequencies ($f_1$ and $f_2$) to satisfy conditions 5, 6 and 7 of \cref{table:collision_types}, as shown in \cref{fig:lattices_and_freqs}. A third type of lattice, the `heavy hexagon', uses a similar scheme. Here the bulk of the lattice includes degree-three and degree-two vertices. Additional degree-two and degree-one vertices lie at the edges. In this lattice, all of the `frequency collisions' and forbidden regions can be satisfied using only three frequencies $f_3 > f_2 > f_1$, with all control qubits residing on degree-two vertices with frequency $f_3$.\\

\begin{figure}
	\centering
	\includegraphics[width=0.48\textwidth, trim=0.0cm 0.0cm 0.0cm 0.0cm, clip]{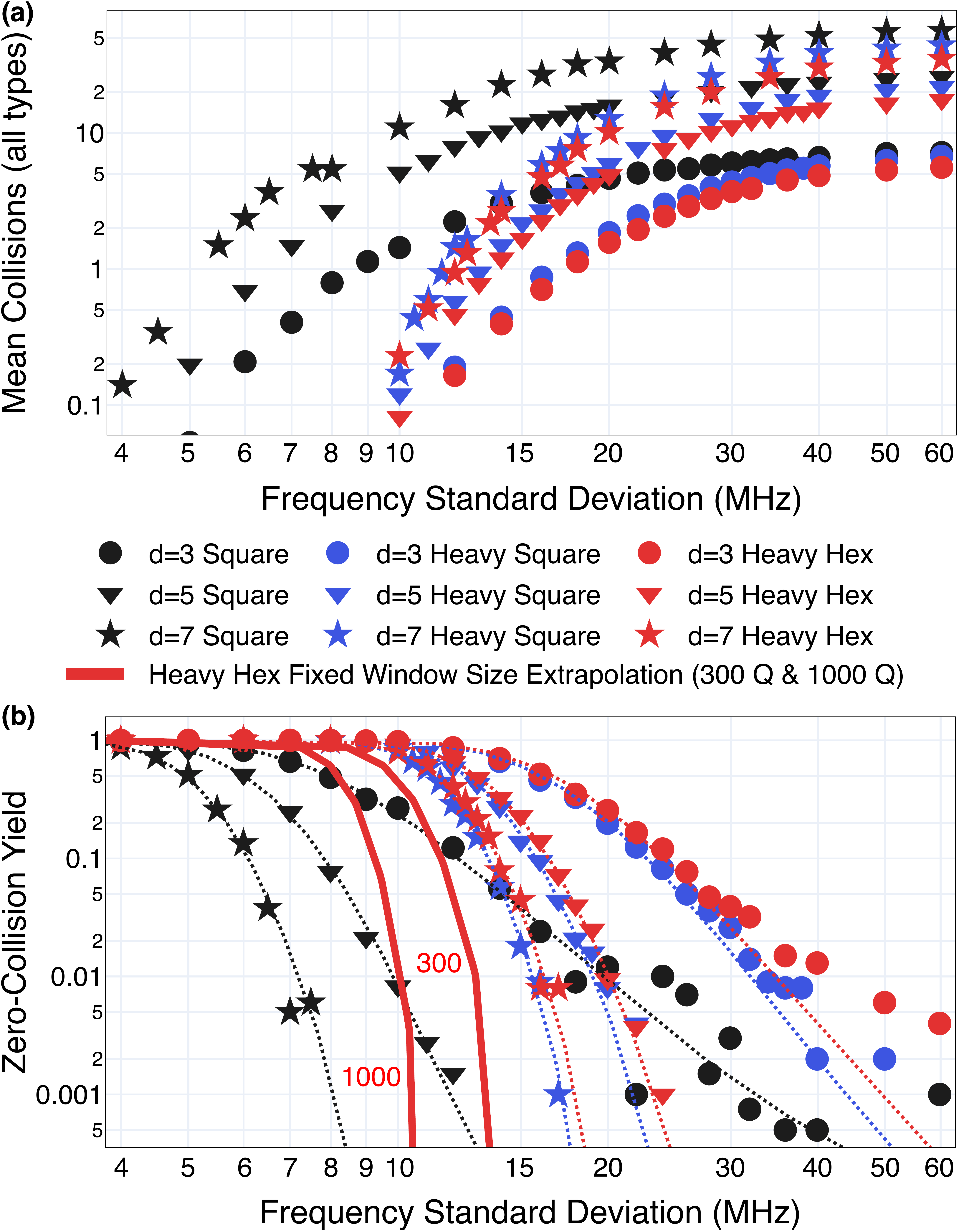}
	\caption{\textbf{Frequency-crowding trends as a function of $\sigma_f$.} Results of Monte Carlo simulation. (See \cref{sec:methods}.) \textbf{(a) Average number of collisions.} and \textbf{(b) Fraction of cases having zero `frequency collisions'.} Simulation was applied to the lattices and frequency patterns shown in \cref{fig:lattices_and_freqs}, capable of distance-5 codes, as well as to $d = 3$ and $d = 7$ scale lattices of square, heavy-square or heavy-hexagon type. (See Supplementary Figures 1-9 for lattice layouts and \cref{table:MC_results} for numbers of qubits.) Color-coded dotted lines in part (b) are fits of each lattice yield to expression $\left[ \int_{-\infty}^{(\Delta f / \sigma_f)} e^{-\frac{1}{2} x^2} dx \right]^N$, where $N$ is the number of qubits and $\pm \Delta f$ defines an allowable `window' around frequency set-points. (See \cref{table:MC_results}.) Using this expression, two solid red lines predict yield for the heavy-hexagon lattice type at 300 and 1000 qubits, using $\Delta f = $ 27.99 and 26.32 MHz, respectively. See \cref{fig:fixedwindowmodel} for estimation of $\Delta f$ as a function of qubit number.} \label{fig:meancollis_vs_sigmaf} \label{fig:yield_vs_sigmaf}
\end{figure}

We use a Monte Carlo model to quantify the frequency-crowding in each lattice type. We sample the qubits at random frequencies drawn from normal distributions characterized by $\sigma_f$, and count the collisions defined in \cref{table:collision_types}. (See \cref{sec:methods}.) In \cref{fig:meancollis_vs_sigmaf}, we show the mean number of `frequency collisions' predicted by the Monte Carlo model for each lattice type and frequency pattern, as a function of $\sigma_f$. As $\sigma_f \rightarrow 0$, the lattice approaches the ideal patterns of  \cref{fig:lattices_and_freqs}, and has zero `frequency collisions'. As $\sigma_f$ increases, the number of  `frequency collisions' rises steadily. As $\sigma_f \rightarrow f_{01} - f_{12}$, the different conditions appearing in \cref{table:collision_types} all become likely, and a limiting number of `frequency collisions' is reached. Yield follows the inverse trend, as seen in \cref{fig:yield_vs_sigmaf}. As $\sigma_f$ increases, the likelihood of finding a `collision free' chip falls off sharply. While the step sizes between frequencies $f_1$ to $f_5$ are important, absolute frequency values are not. Setting $f_1= 5.0$, $f_2 = 5.07$ and $f_3 = 5.14$ GHz works as well as $f_1= 5.05$, $f_2 = 5.12$ and $f_3 = 5.19$ GHz. \\

The yield and mean collision number are a function of the several different collision types and bounds, so they are not readily susceptible to an analytic formulation. However, we can propose a simplified model for yield: in order for a lattice to be collision-free, every qubit in the lattice must fall within some frequency `window' $\pm \Delta f$ relative to its setpoint. Presuming the qubit frequencies are normally distributed, the probablity of this occurring goes as the cumulative distribution function, raised to the power $N$, where $N$ is the number of qubits: $\left[ \int_{-\infty}^{(\Delta f / \sigma_f)} e^{-\frac{1}{2} x^2} dx \right]^N$. In the yield plot in \cref{fig:yield_vs_sigmaf}, we fit this expression to find $\Delta f$ for each lattice. \\

These model results allow us to predict how different lattice types and frequency patterns will respond to fabrication imprecision. As shown in \cref{fig:meancollis_vs_sigmaf}, if imprecision $\sigma_f$ is greater than 30 MHz, any $d = 5$ lattice will exhibit $>10$ `frequency collisions' of one or another of the types listed in \cref{table:collision_types}, causing the affected gates to have error rates above $\sim$ 1 \%. However, if $\sigma_f = 10$ MHz then on average the $d = 5$ square lattice will exhibit 5 `frequency collisions', while the `heavy square' and `heavy hexagon' lattices will exhibit 0.1 `frequency collision'. Considered in terms of yield, we see from \cref{fig:yield_vs_sigmaf} that if $\sigma_f = 10$ MHz, then for a $d = 5$ device, a square lattice with 5-frequency pattern has a 0.8\% likelihood to be `collision free', whereas a `heavy square' lattice with 3-frequency pattern has 90\% likelihood and `heavy hexagon' with 3-frequency pattern has 92\% likelihood. Alternatively we can ask, how well do we have to control $\sigma_f$? If we seek a 10\% yield, then \cref{fig:yield_vs_sigmaf} indicates that for a $d = 5$ device, a square lattice with 5-frequency pattern requires $\sigma_f < 8$ MHz, whereas a `heavy square' lattice with 3-frequency pattern requires $\sigma_f = 16$ MHz and `heavy hexagon' with 3-frequency pattern requires $\sigma_f = 17$ MHz. Although the square lattice requires 10 to 20 \% fewer qubits than the other types at each distance $d$, it requires far better frequency precision.\\

The as-fabricated $\sigma_f$ seen in \cref{fig:36Q_andTuned2f} is 132.3 MHz. (See \cref{sec:results}) The Monte Carlo modeling finds that for a heavy-hexagon lattice at $d=3$ scale this $\sigma_f$ can enable 0.1\% yield of collision-free chips. Other lattice types and larger scales will all have yield $\ll 0.1\%$. The re-tuned $\sigma_f$ = 14.0 MHz demonstrated in \cref{fig:36Q_andTuned2f} will improve the yield in all types of lattice. Predictions of the Monte Carlo model for $\sigma_f$ = 14.0 MHz appear in \cref{table:MC_results}. At $d = 5$ scale the heavy-hexagon and heavy-square lattices and 3-frequency patterns should be collision-free nearly one-third of the time, while at $d = 7$ scale the yield is about four times smaller, still reasonable for prototype systems. \\

\begin{figure}
	\centering
	\includegraphics[width=0.48\textwidth, trim=0cm 0cm 0cm 0cm, clip]{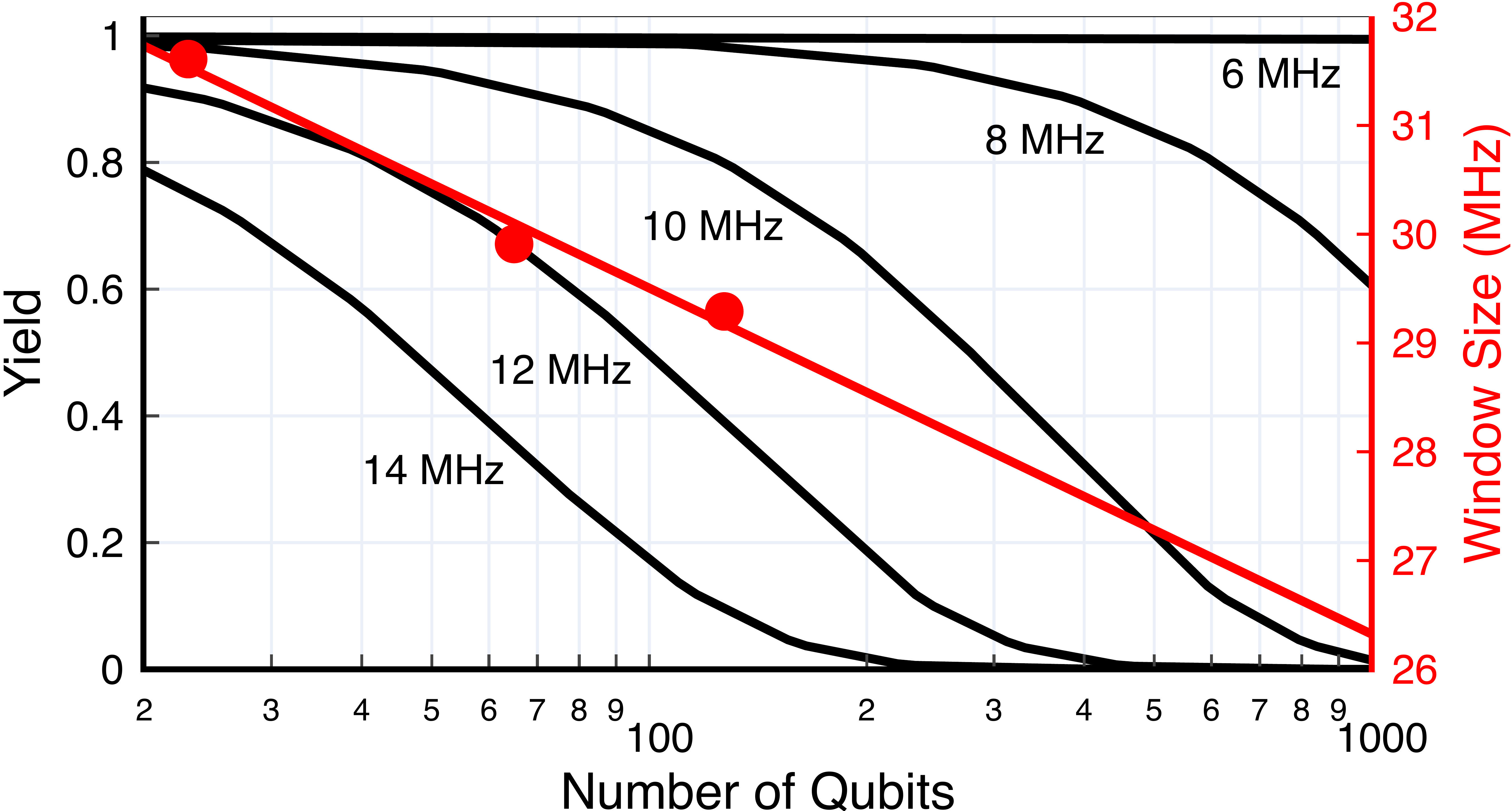}\\
	\caption{\textbf{Yield scaling for heavy hexagon lattice.} Fixed window size model. (left axis) Model yields for $\sigma_f$ ranging from the 14 MHz demonstrated in this work to 6 MHz that would optimize future scaling yield beyond the 1000Q level. (right axis) Fixed window size fits to the d=3, d=5 and d=7 heavy hexagon lattice, as well as a fit of these values to expression $A + B \cdot \log(N_{QB}) $, extrapolated from 20 to 1000 qubits. This trend illustrates varying frequency crowding constraints as a function of lattice size. } \label{fig:fixedwindowmodel} \end{figure}

As seen from the Monte Carlo analysis, the laser-anneal rework method can scale to the $>$ 100 qubit size, enabling a well-chosen lattice and frequency pattern to implement $d = 7$ error-correction codes free of frequency-crowding. To examine needs for the next generation of chips up to the 1000-qubit level, we can coarsely estimate requirements by extrapolating the fixed window model for the heavy-hexagon lattice as shown in \cref{fig:fixedwindowmodel}. While the $\sigma_f$ = 14.0 MHz demonstrated here enables practical yield up to the 100 to 200 qubit scale, it is clear that roughly a factor of two further improvement is needed to scale towards 1000 qubits. Since this precision is also better than the resistance-to-frequency prediction precision shown in this work, development of further refinements in tuning and frequency prediction approaches will be necessary as the scale of fixed-frequency transmon circuits surpass the 100 qubit milestone. 

\begin{table*}[htp]
	\centering
        \begin{tabular}{| >{\centering}p{0.1\textwidth} || >{\centering}p{0.1\textwidth} | >{\centering}p{0.1\textwidth} || >{\centering}p{0.1\textwidth} | >{\centering}p{0.1\textwidth} || >{\centering}p{0.1\textwidth} | >{\centering}p{0.1\textwidth} || >{\centering\arraybackslash}p{0.1\textwidth} |}
        	    \hline
	    			&		&		& \multicolumn{2}{c||}{$\sigma_f = 132.3$ MHz} 	& \multicolumn{2}{c||}{$\sigma_f = 14$ MHz}&			\\
	    \cline{4-7}
            Lattice \& 	& Code 	& Qubits	& Mean 		& Yield of 		& Mean 		& Yield of 			& $\Delta f$ 	\\
            frequency 	& distance	& 	 	& number	of	& collision-	& number	of	& collision-		& Window		\\
            pattern 		&	 	&  		& collisions 	& free devices	& collisions 	& free devices 		& (MHz)		\\
            \hline 
            Square, 	& $d = 3$ & 17 		& 9 			&  $\ll 0.1\%$ 	& 3			& 6\%			& 13.96 \\
            5-frequency 	& $d = 5$ & 49 		& 35 			& $\ll 0.1\%$ 	& 10 			& $<0.1\%$ 		&13.23 \\
             			& $d = 7$ & 97 		& 78		& $\ll 0.1\%$ 	& 23			& $\ll 0.1\%$ 		&12.12 \\
            \hline
            Heavy 		& $d = 3$ & 25 		& 10 		& $\ll 0.1\%$ 		& 0.4		& 67\%  		& 30.89 \\
            square, 3-	& $d = 5$ & 73 		& 33		& $\ll 0.1\%$ 		& 1.5		& 27\%  		& 29.49 \\
            frequency 	& $d = 7$ & 145 	& 70 		& $\ll 0.1\%$ 		& 3.5		& 6\%  		& 29.06\\
            \hline
            Heavy 		& $d = 3$ & 23 		& 8 		& $0.1\%$ 		& 0.4			& 70\%  		& 31.61 \\
            hexagon, 	& $d = 5$ & 65 		& 25 		& $\ll 0.1\%$ 		& 1.2		& 33\%  		& 29.91 \\
            3-frequency 	& $d = 7$ & 127 	& 51	 	& $\ll 0.1\%$ 		& 2.7		& 8\%  		& 29.29 \\
            \hline
        \end{tabular}
	\caption{\label{table:MC_results} \textbf{Monte-Carlo model predictions of mean number of collisions and yield of collision-free chips.} Full results of Monte-Carlo model of frequency-crowding appear in \cref{fig:meancollis_vs_sigmaf}. Here we show values for as-fabricated precision $\sigma_f = 132.3$ MHz, and for demonstrated frequency precision of the laser-anneal method, $\sigma_f = 14$ MHz. (\cref{fig:36Q_andTuned2f}) in different lattices and frequency patterns (\cref{fig:lattices_and_freqs} and Supplementary Figures 1-9). The $\Delta f$ values correspond to the fit lines in \cref{fig:yield_vs_sigmaf}: in order for the device to be collision-free, every qubit in the lattice must fall within $\pm$ this `window', relative to its set-point.}
\end{table*}

\section{Methods} \label{sec:methods}

\subsection{Chip Fabrication}

A chip of the kind used to determine $\sigma_f$ and to test our laser-anneal rework process appears in \cref{fig:36Q_andTuned2f}. All microwave elements comprise Nb films $\sim$ 200 nm thick on a silicon substrate. Each qubit is coupled to a readout resonator but is not directly coupled to any nearby qubits. All transmon capacitors are identical. Junctions are fabricated using identical electron-beam lithographic patterns and deposited simultaneously using double-angle deposition and oxidation \cite{dolan_offset_1977}. The individual qubit design is similar to that used in Ref \cite{sheldon_procedure_2016} with aharmonicity $f_{12} - f_{01} \simeq $ -330 MHz. Junctions have linear dimension $\sim 100$ nm and are designed for $I_c$ of $\sim$ 30 nA. During packaging, we accidentally damaged three of the 36 qubits and found these to be non-functional when cooled in a dilution refrigerator. We left two of the remaining 33 qubits un-tuned as experimental controls, so that our tuning demonstration includes 31 qubits. \\

\subsection{Tuning Using Selective Laser Anneal}

We have built an integrated junction rework system that can measure and modify the junction resistance. Fig. \ref{fig:LaserTuneSchematic} shows a schematic of our laser annealing system, which we call ``Laser Annealing of Stochastically Impaired Qubits'' (LASIQ). The laser output is generated by a diode-pumped solid-state laser, frequency doubled to 532 nm. Active power control of the anneal beam is performed using a piezo-rotary mounted waveplate and polarizing beam splitter (PBS), which is adaptively adjusted based on a pick-off beam measured on a downstream silicon photodiode. A precision-timed shutter is used to control the anneal duration, and beam alignment is performed using a mechanical mirror mount which directs the beam via pattern recognition to the transmon junction center. The beam is shaped as needed to avoid illuminating the junction directly \cite{rosenblatt_laser_2019-1}. \\

By careful control of laser power and pulse duration, we use this system to adjust $R_n$. This process overcomes the imprecision due to transmon fabrication, with a residual imprecision $\sigma_f$ due to the rework process. To develop the process, we prepared a set of more than 150 junctions identically to qubit junctions, and measured their response to a range of laser powers and exposure times. We recorded $R_n$ shifts up to 15\% relative to initial $R_n$, for anneal durations varying by an order of magnitude and laser powers varying by 20\%. Response to laser power in particular was highly nonlinear. Based on these empirical calibrations of $R_n$ shift to power and exposure, we established a qubit tuning process: We first measure the transmon junction's $R_n$ using four-point probing of the transmon capacitor pads at 25 \si{\degree}C. Using a $f(R_n)$ prediction based on a previously determined correlation curve (\cref{fig:36Q_andTuned2f}), we assign the junction a target resistance corresponding to the target frequency in a multiqubit chip lattice. Because the anneal can shift $R_n$ in only one direction, the target must be higher than the initial $R_n$. We anneal the qubit junction using laser power and duration chosen from our calibration set, then re-measure its $R_n$. A junction requiring large shifts in $R_n$ may require repeated anneals to reach its target, as shown in \cref{fig:LaserTuneSchematic}. The control algorithm increases the resistance until the measured value is within 0.3\% of the target value. In a separate trial of tuning precision, more than 300 junctions were tuned to target $R_n$s ranging from 0.4\% to 14.5\% above their initial values, and landed successfully within this 0.3\% margin. We expect 0.3\% imprecision in $R_n$ to introduce 0.15\% imprecision in transmon frequency.\\

\subsection{Monte Carlo Frequency-Crowding Model}

\begin{figure}
	\centerline{\includegraphics[width=9cm, trim=0.0cm 7.0cm 9.0cm 0.0cm, clip]{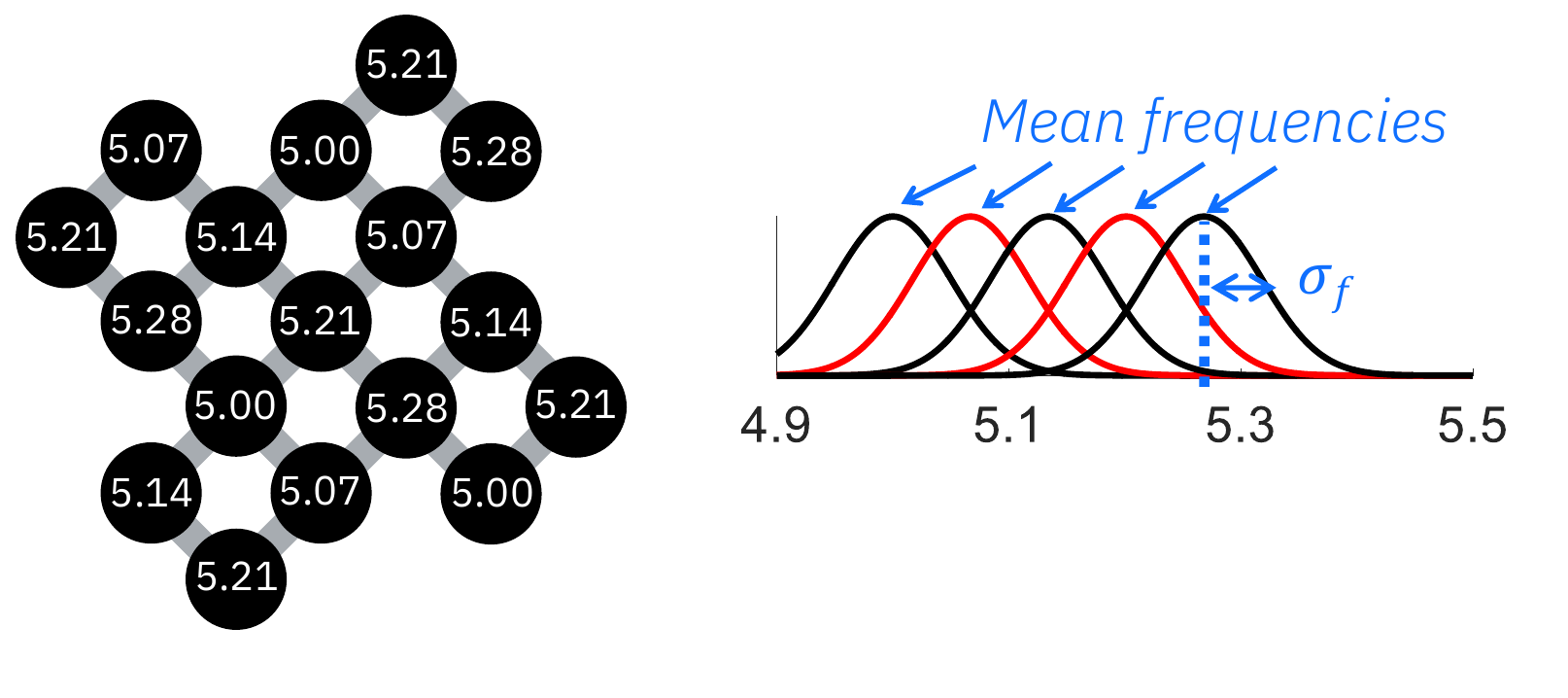}}
	\caption{\textbf{Frequency collision statistical model.} \textbf{Left: } Square lattice, $d = 3$ with 5-frequency pattern of \cref{fig:lattices_and_freqs}: $f_1 = 5.00$, $f_2 = 5.07$, $f_3 = 5.14$, $f_4 = 5.21$, $f_5 = 5.28$ GHz. To model the lattice statistically, treat the frequencies $f_1$ to $f_5$ as means of distributions. \textbf{Right: } Mean frequencies and normal distributions characterized by $\sigma_f$. For each position in the lattice, sample from the local distribution. Choose random frequencies in this fashion, count collisions as described in \cref{table:collision_types} and repeat to gather statistical sample. Process is repeated for differing spacings between mean frequencies $f_1$ to $f_5$, and for different distribution widths $\sigma_f$.}
	\label{fig:freqpatternexample}
\end{figure}

Using a Monte Carlo model, we can estimate the incidence of `frequency collisions' in a lattice as a function of $\sigma_f$. We assume that imperfect frequency-setting will distribute qubit frequencies normally around their design frequencies with standard deviation $\sigma_f$. For lattices of the type shown in \cref{fig:lattices_and_freqs}, we designate 3 to 5 frequencies $f_1$, $f_2$, $f_3$, $f_4$, $f_5$ spaced at regular intervals in the pattern shown. We set $f_1 = 5$ GHz, similar to real-world transmons \cite{sheldon_characterizing_2016, jurcevic_demonstration_2020}. We sample the qubit frequencies randomly around these values and count the collisions throughout the lattice, as listed in \cref{table:collision_types}. This process is illustrated in \cref{fig:freqpatternexample}. We repeat the frequency-assignment and counting to build statistics for a given lattice and frequency pattern. We then repeat the model for a range of $\sigma_f$ values from 0 to 150 MHz. We repeat the entire process over a range of frequency spacings, to find the spacing that minimizes `frequency collisions' at each value of $\sigma_f$. As a function of $\sigma_f$ we can then extract 1) the mean number of total collisions in the lattice, and 2) the fraction of repetitions which result in zero collisions (`yield'). Our simulations used 1000 repetitions except to find yield below 1\% in $d = 5$ lattices and below 0.2\% in $d = 3$ lattices, which used 4000 repetitions, and in $d = 7$ lattices to find mean collisions for $\sigma_f < 16$ MHz or yield above 50\% (100 repetitions) or to find mean collisions for $\sigma_f > 16$ MHz (40 repetitions). \\ 
 
\newpage
\bibliographystyle{naturemag}
\bibliography{Yield_and_Tuning_Bib}

\section*{Acknowledgements}
We acknowledge funding from the Intelligence Advanced Research Projects Activity (IARPA) under contract W911NF-16-1-0114, for the multi-qubit test vehicle and frequency-vs-resistance correlation studies. We thank N. Bronn, M. Carroll, C. Chamberland, A. Cross, J. Gambetta, J. Ku, M. Malekakhlagh, D. McKay, B. Plourde, E. Pritchett, A. Rosenbluth, M. Takita, J. Timmerwilke and G. Zhu for helpful discussions. We thank E. Porter for coding assistance, Y. Martin and R. Haight for assistance in constructing the laser optics, and R. Patel for photomicroscopy.

\end{document}


\captionsetup[figure]{labelfont={bf},name={Supplementary Figure},labelsep=period}

\title{SUPPLEMENTARY INFORMATION for \\Laser annealing Josephson junctions for yielding scaled-up superconducting quantum processors)}

\author{Jared B. Hertzberg}
\author{Eric J. Zhang}
\author{Sami Rosenblatt}
\author{Easwar Magesan}
\author{John A. Smolin}
\author{Jeng-Bang Yau}
\author{Vivekananda P. Adiga}
\author{Martin Sandberg}
\author{Markus Brink}
\author{Jerry M. Chow}
\author{Jason S. Orcutt}
\affiliation{IBM Quantum, IBM T.J. Watson Research Center, Yorktown Heights, NY 10598, USA}

\maketitle{}


\begin{figure}
	\includegraphics[width=0.3\textwidth, trim=0.0cm 0.0cm 0.0cm 0.0cm, clip]{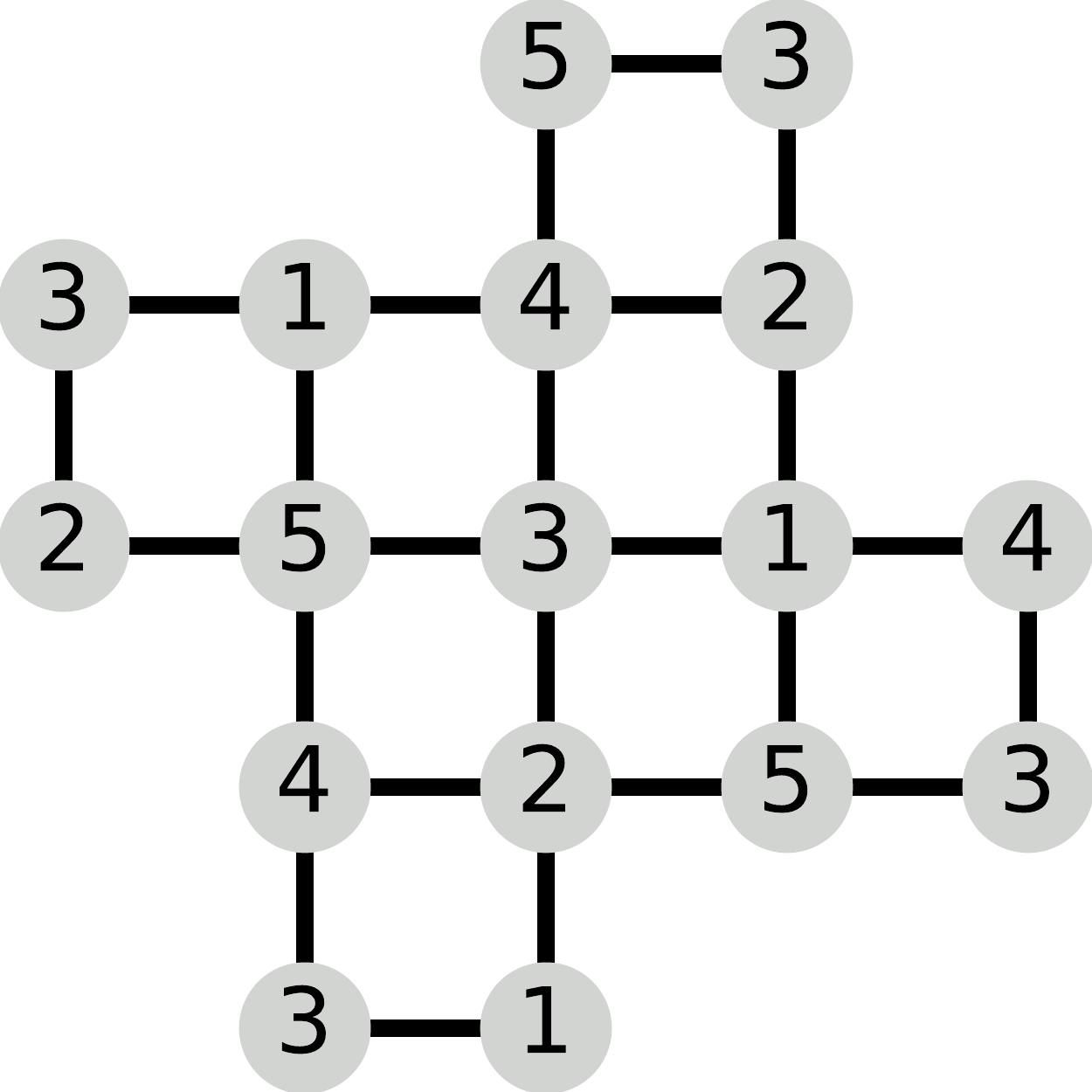}
	\caption{Distance-3 square lattice, 17 qubits, 5-frequency pattern} \label{}
\end{figure}

\begin{figure}
	\includegraphics[width=0.4\textwidth, trim=0.0cm 0.0cm 0.0cm 0.0cm, clip]{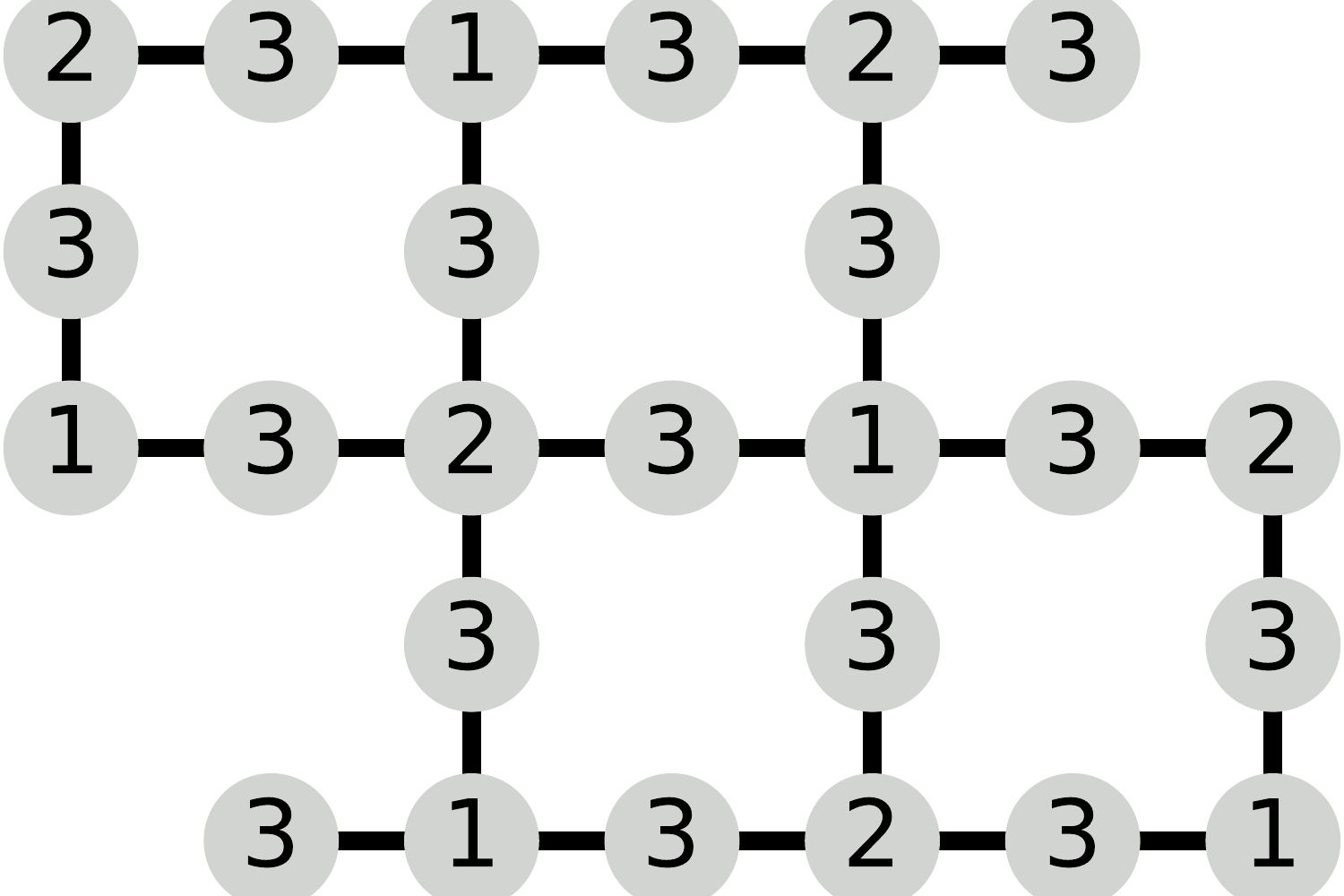}
	\caption{Distance-3 heavy-square lattice, 25 qubits, 3-frequency pattern} \label{}
\end{figure}

\begin{figure}
	\includegraphics[width=0.4\textwidth, trim=0.0cm 0.0cm 0.0cm 0.0cm, clip]{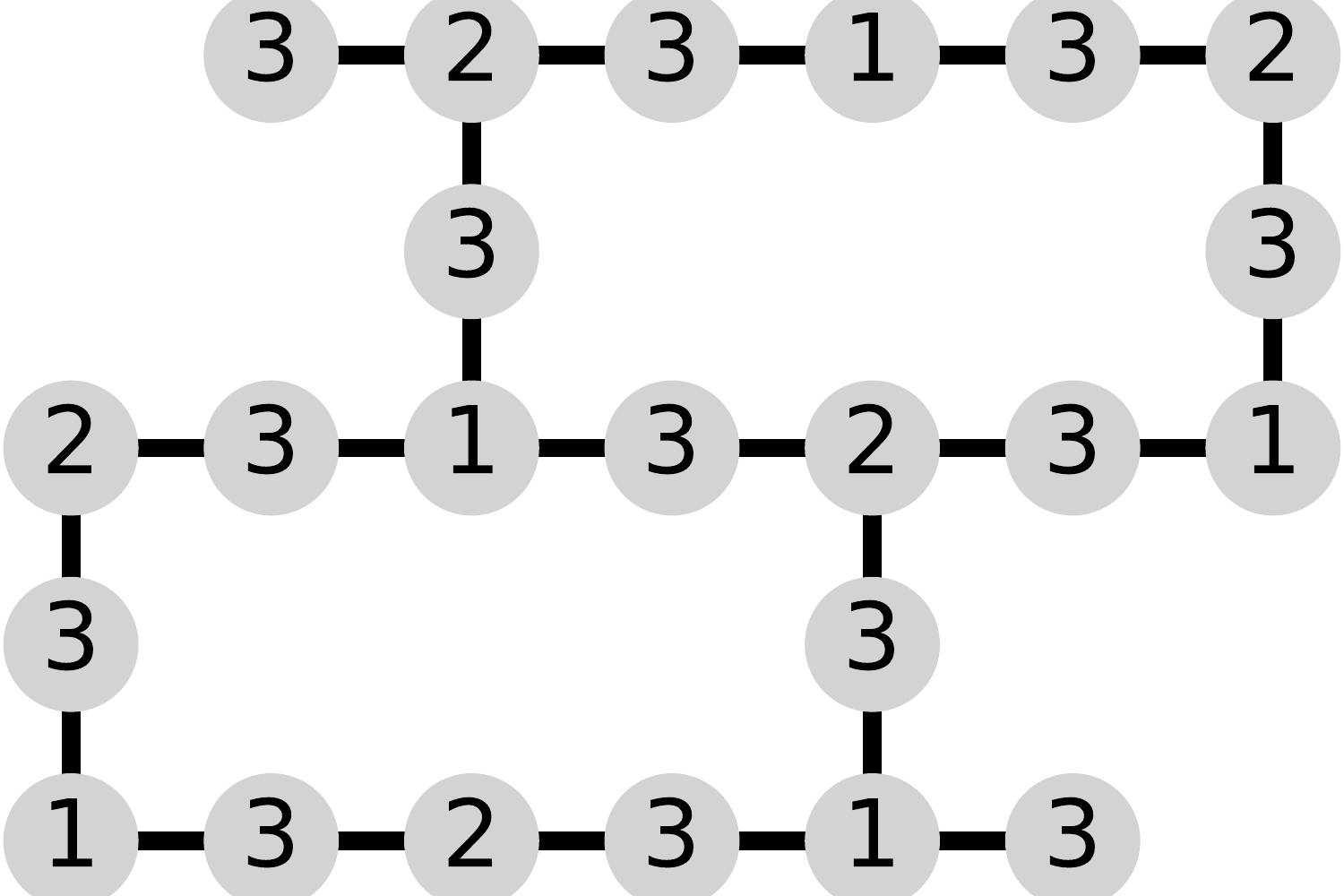}
	\caption{Distance-3 heavy-hexagon lattice, 23 qubits, 3-frequency pattern} \label{}
\end{figure}

\begin{figure}
	\includegraphics[width=0.55\textwidth, trim=0.0cm 0.0cm 0.0cm 0.0cm, clip]{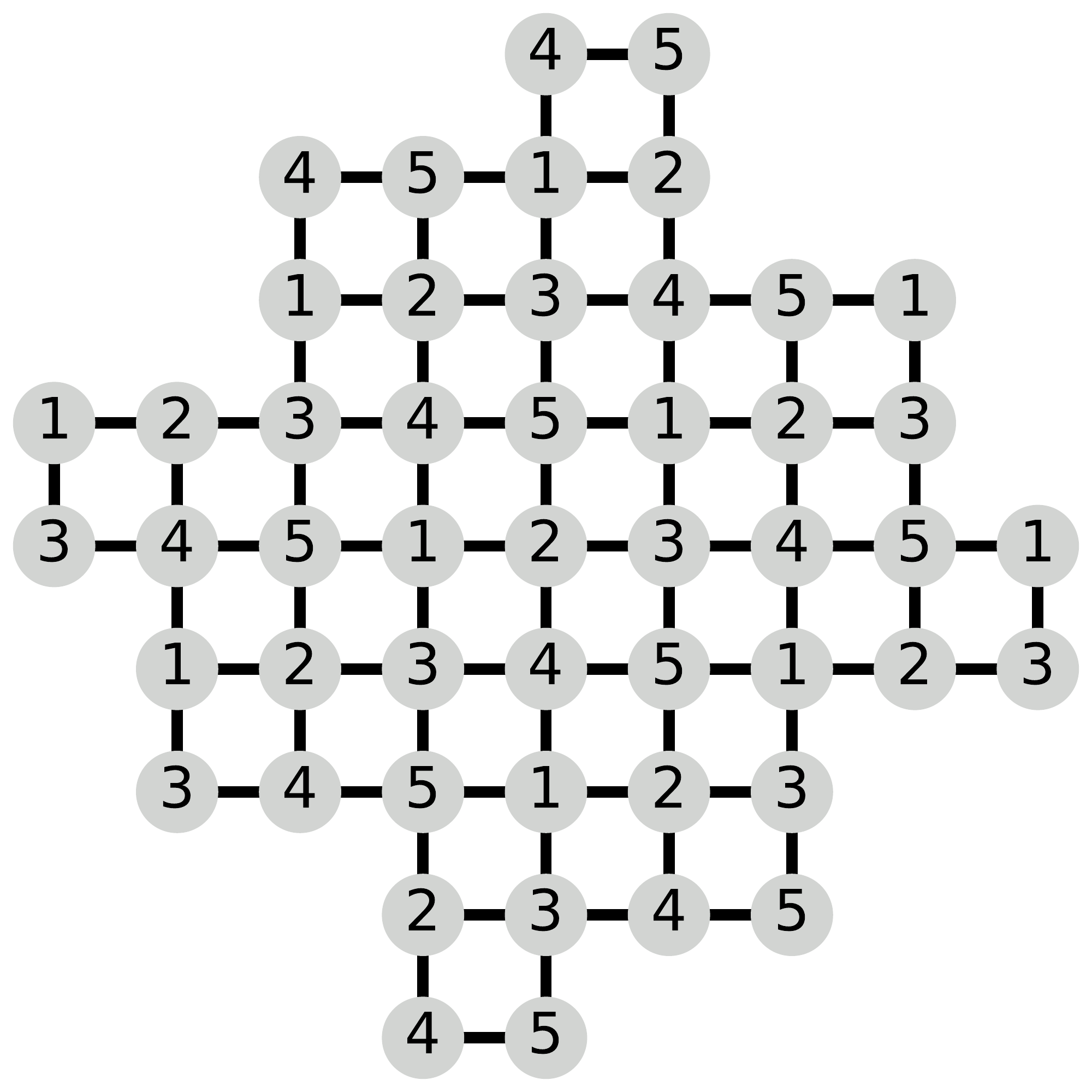}
	\caption{Distance-5 square lattice, 49 qubits, 5-frequency pattern} \label{}
\end{figure}

\begin{figure}
	\includegraphics[width=0.65\textwidth, trim=0.0cm 0.0cm 0.0cm 0.0cm, clip]{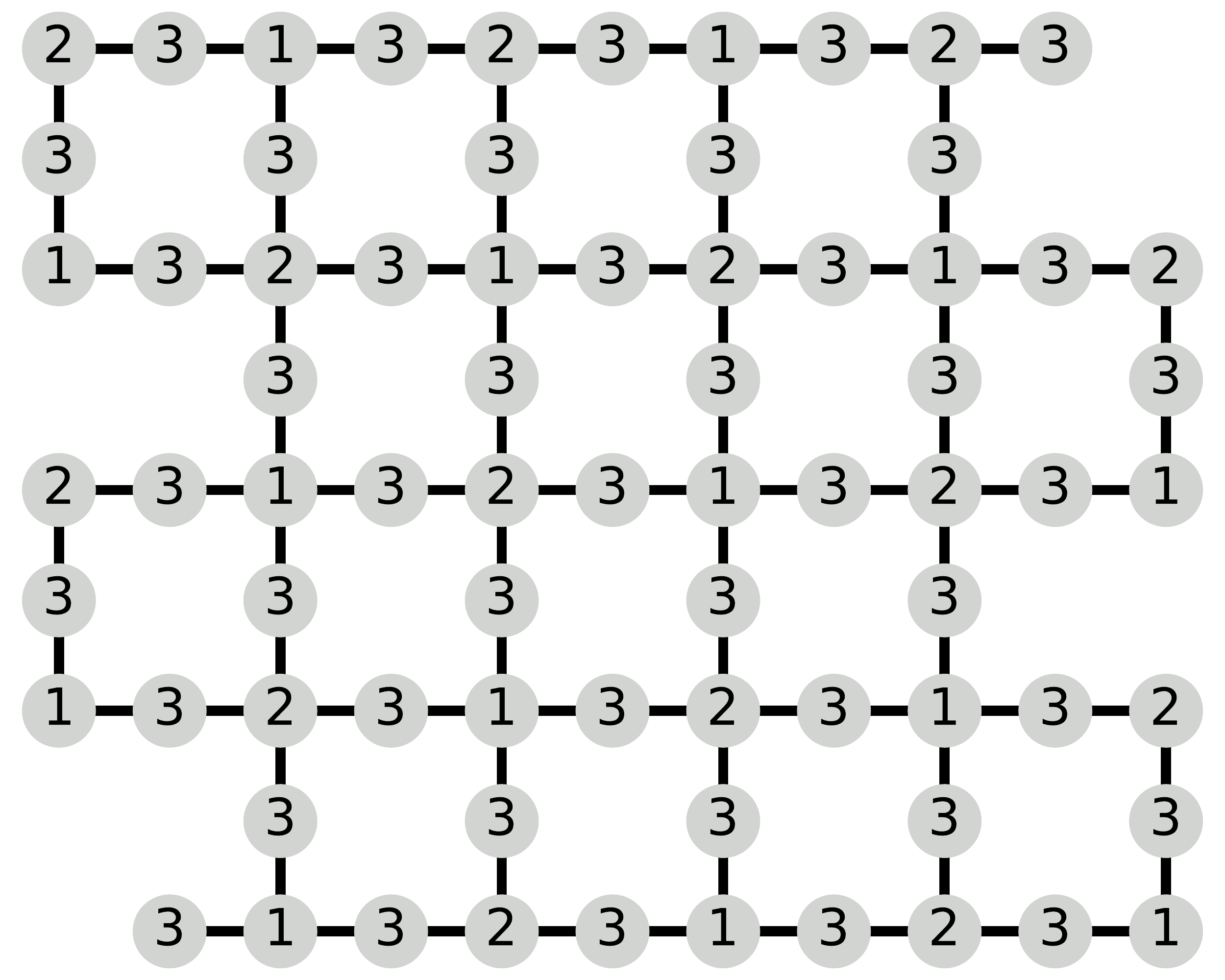}
	\caption{Distance-5 heavy-square lattice, 73 qubits, 3-frequency pattern} \label{}
\end{figure}

\begin{figure}
	\includegraphics[width=0.6\textwidth, trim=0.0cm 0.0cm 0.0cm 0.0cm, clip]{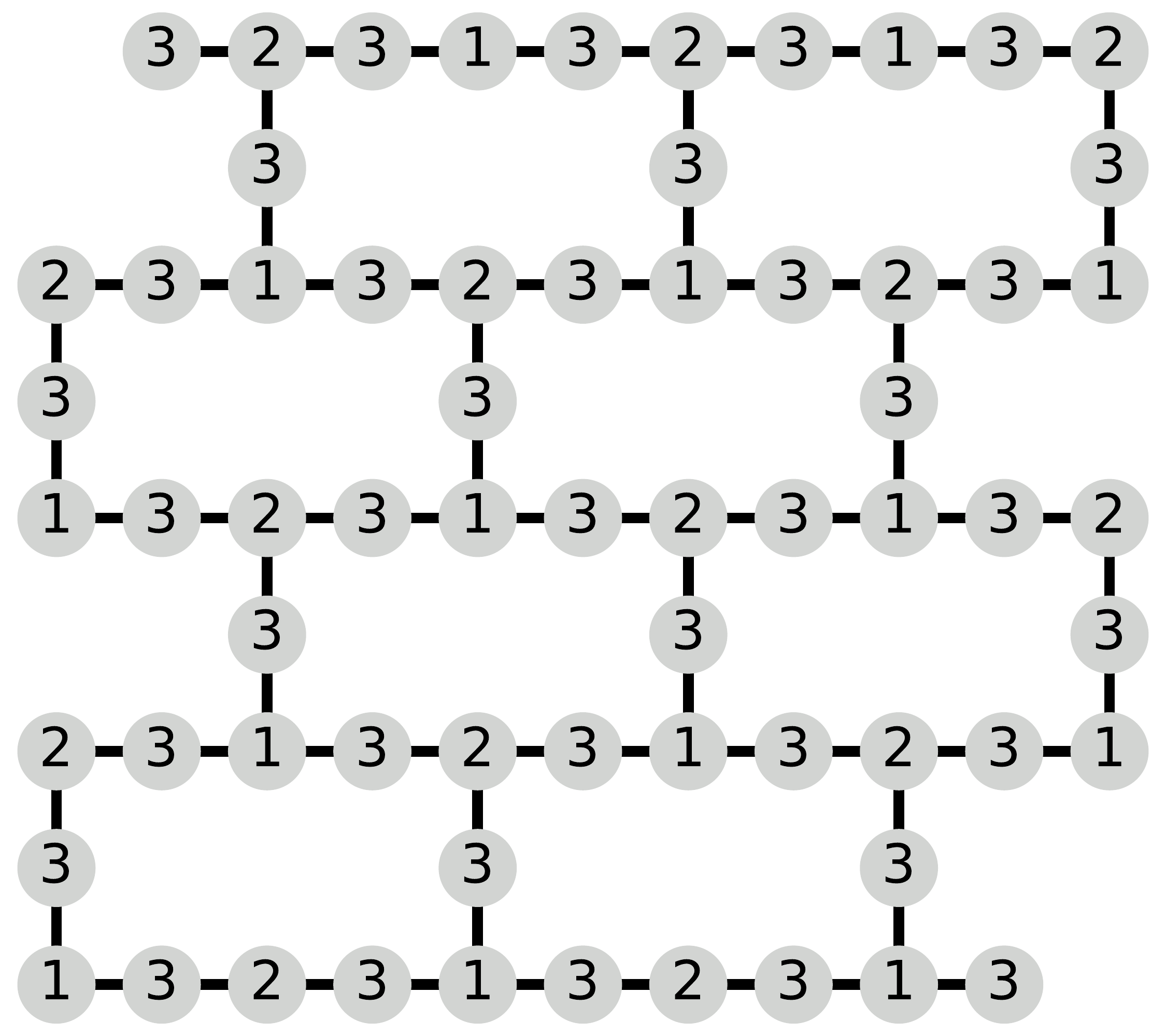}
	\caption{Distance-5 heavy-hexagon lattice, 65 qubits, 3-frequency pattern} \label{}
\end{figure}

\begin{figure}
	\includegraphics[width=0.8\textwidth, trim=0.0cm 0.0cm 0.0cm 0.0cm, clip]{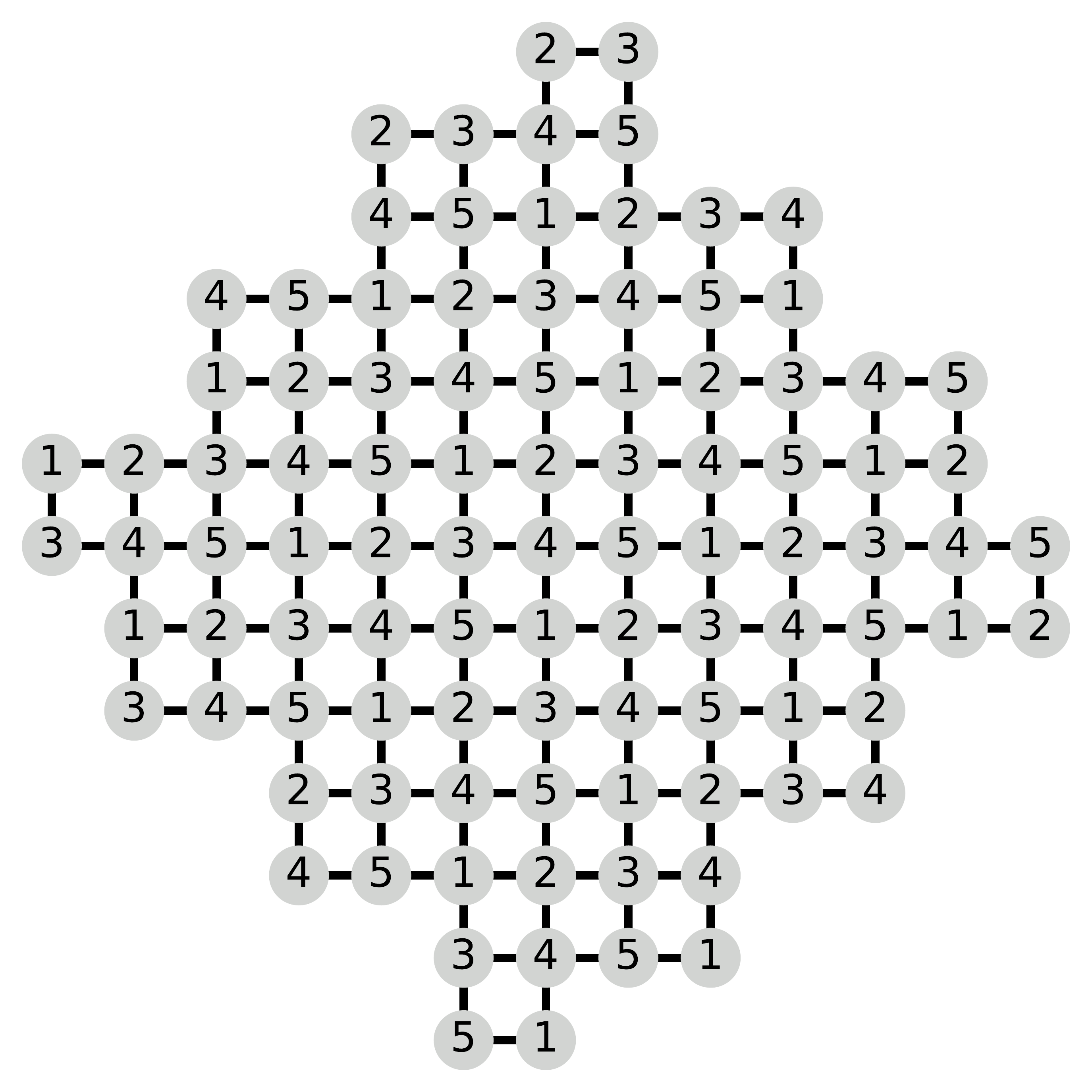}
	\caption{Distance-7 square lattice, 97 qubits, 5-frequency pattern} \label{}
\end{figure}

\begin{figure}
	\includegraphics[width=0.9\textwidth, trim=0.0cm 0.0cm 0.0cm 0.0cm, clip]{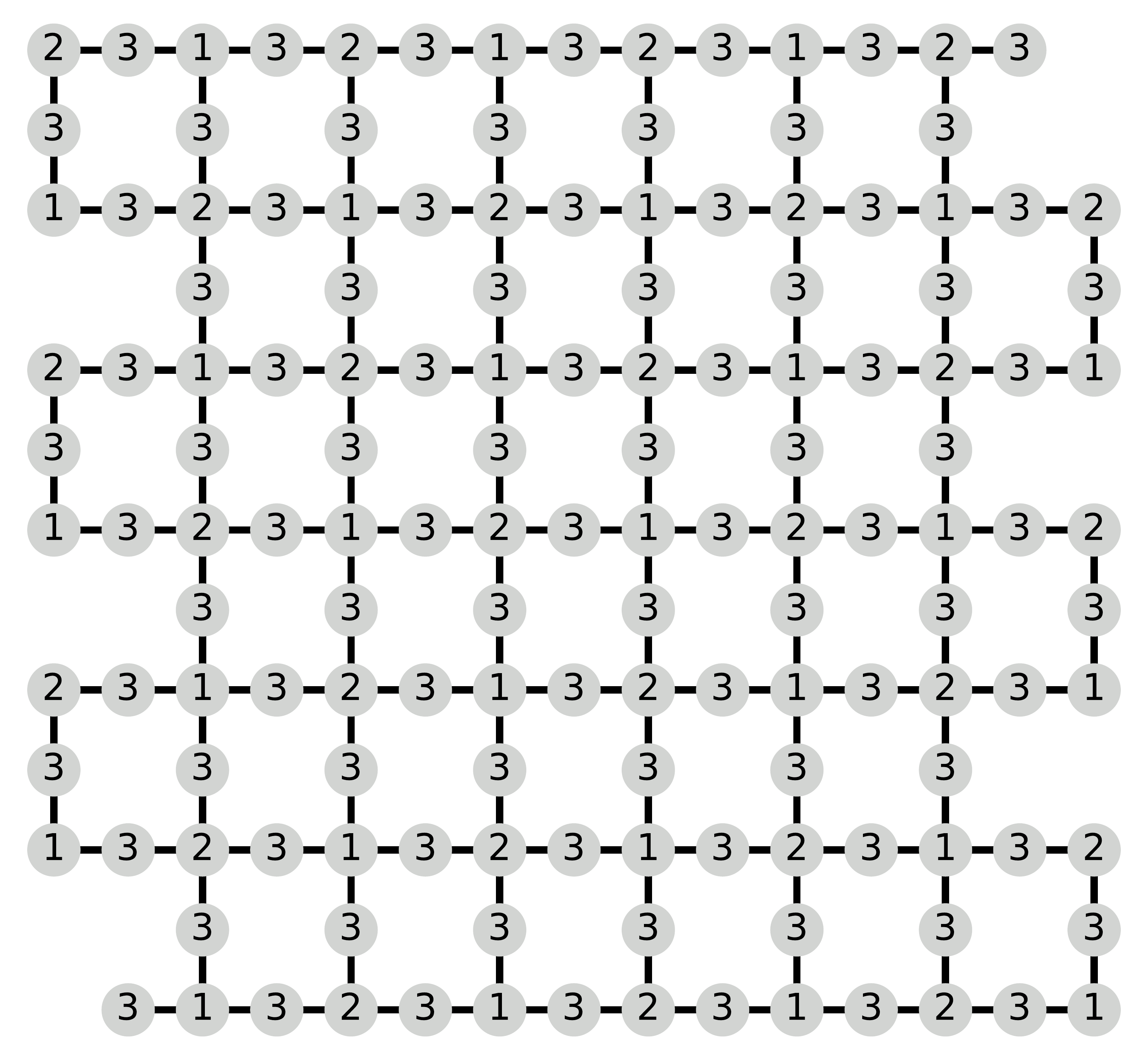}
	\caption{Distance-7 heavy-square lattice, 145 qubits, 3-frequency pattern} \label{}
\end{figure}

\begin{figure}
	\includegraphics[width=0.95\textwidth, trim=0.0cm 0.0cm 0.0cm 0.0cm, clip]{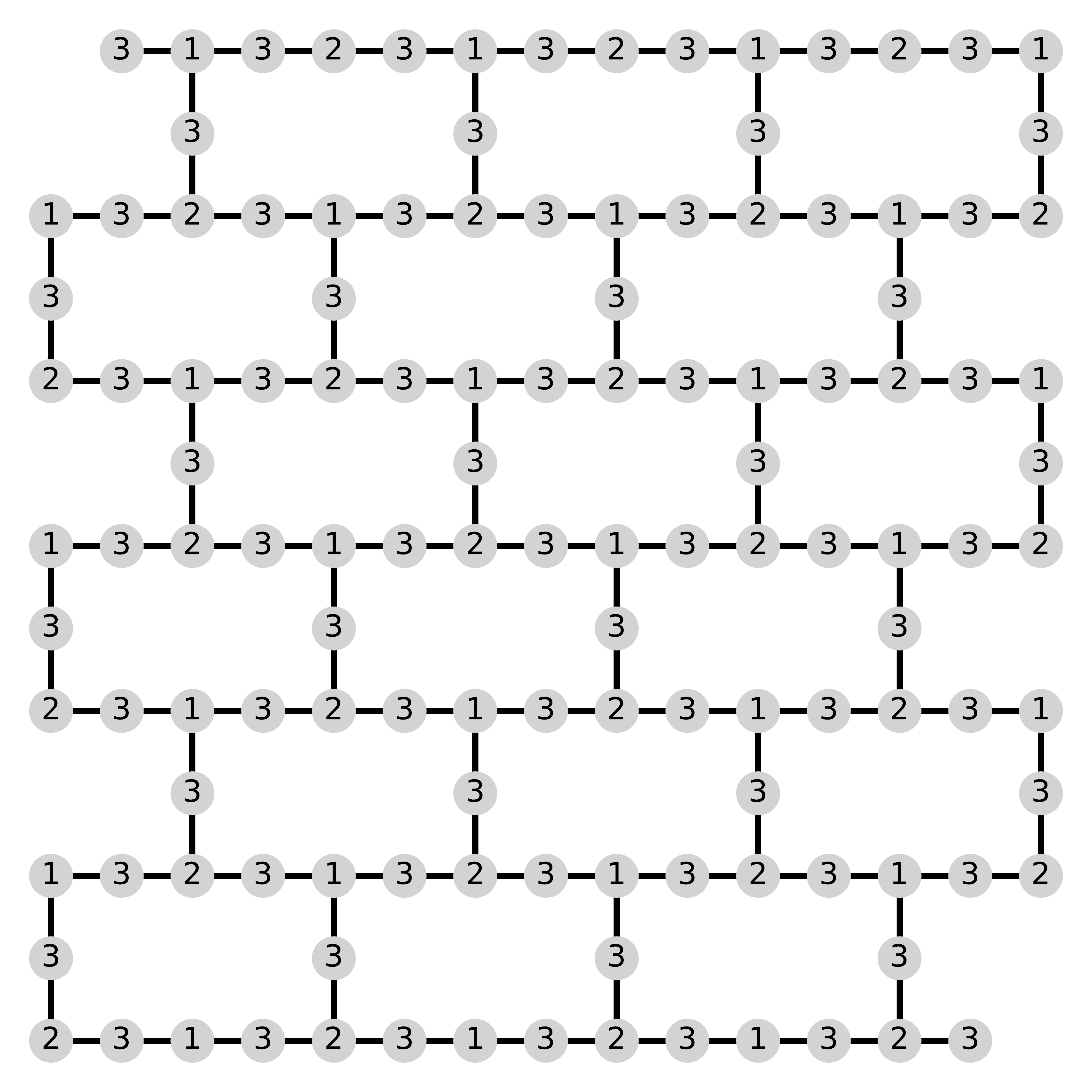}
	\caption{Distance-7 heavy-hexagon lattice, 127 qubits, 3-frequency pattern} \label{}
\end{figure}